\documentclass[a4paper,11pt]{article}
\pdfoutput=1 

\usepackage{jcappub} 

\usepackage[T1]{fontenc} 
\usepackage{natbib}
\usepackage{cleveref}

\usepackage{subfig}

\usepackage{siunitx}
\DeclareSIUnit\Mpc{Mpc}
\DeclareSIUnit\Gpc{Gpc}
\DeclareSIUnit\Gyr{Gyr}

\usepackage{bm}

\newcommand{\FastDF}{{\sc{fastdf}}}
\newcommand{\CLASS}{{\sc{class}}}
\newcommand{\monofonIC}{{\sc{monofonic}}}

\title{\boldmath Geodesic motion and phase-space evolution of massive neutrinos}

\author[a]{Willem Elbers}

\affiliation[a]{Institute for Computational Cosmology, Department of Physics, Durham University, South Road, Durham, DH1 3LE, UK}

\emailAdd{willem.h.elbers@durham.ac.uk}

\abstract{The non-trivial phase-space distribution of relic neutrinos is responsible for the erasure of primordial density perturbations on small scales, which is one of the main cosmological signatures of neutrino mass. In this paper, we present a new code, \FastDF, for generating $1\%$-accurate particle realisations of the neutrino phase-space distribution using relativistic perturbation theory. We use the geodesic equation to derive equations of motion for massive particles moving in a weakly perturbed spacetime and integrate particles accordingly. We demonstrate how to combine geodesic-based initial conditions with the $\delta f$ method to minimise shot noise and clarify the definition of the neutrino momentum, finding that large errors result if the wrong parametrisation is used. Compared to standard Lagrangian methods with ad-hoc thermal motions, \FastDF{} achieves substantial improvements in accuracy. We outline the approximation schemes used to speed up the code and to ensure symplectic integration that preserves phase-space density. Finally, we discuss implications for neutrino particles in cosmological $N$-body simulations. In particular, we argue that particle methods can accurately describe the neutrino distribution from $z=10^9$, when neutrinos are linear and ultra-relativistic, down to $z=0$, when they are nonlinear and non-relativistic. \FastDF{} can be used to set up accurate initial conditions (ICs) for $N$-body simulations and has been integrated into the higher-order IC code \monofonIC.}


\begin{document}
\maketitle
\flushbottom


\section{Introduction}

It is expected that relic neutrinos of the early Universe outnumber the baryons by a factor of $n_\nu/n_b\approx 10^9$. The discovery of neutrino oscillations \cite{fukuda98,ahmad02} implies that at least two-thirds of these particles carry a mass, which though small, through sheer abundance should leave an imprint on the large-scale distribution of matter. Detecting this signature would provide a means of measuring the sum of neutrino masses $\sum m_\nu$ from cosmology \cite{bond80,frenk83,hu98}, complementing an extensive programme of neutrino experiments on Earth. The imprint of massive neutrinos arises primarily from the fact that, during the era of structure formation, neutrinos are non-relativistic particles with a relativistic phase-space distribution. Neutrinos decouple from the primordial plasma at a temperature of $1\,\rm{MeV}$ and subsequently stream along geodesics, essentially without scattering, but maintaining a thermal phase-space distribution. After becoming non-relativistic, massive neutrinos have a thermal velocity $v_{\rm{th}}\propto1/m_\nu$ and cannot be contained effectively in regions smaller than $v_{\rm{th}}/H$, where $H$ is the Hubble rate. As a result, although neutrinos contribute like dust to the geometric expansion of the Universe, they cluster less effectively on small scales, slowing down the growth of matter perturbations. This effect has been used to put tight constraints on the sum of neutrino masses, with current limits of $\sum m_\nu<0.15\,\rm{eV}$ or better \cite{palanque20,choudhury20,DES21,divalentino21}. These constraints are an order of magnitude below the strongest laboratory constraint, $m_\nu < 0.8\,\rm{eV}$, from KATRIN \cite{aker21}, but come with the important assumption of $\Lambda$CDM cosmology, which highlights their complementarity.

Cosmological $N$-body simulations are widely used to make predictions for nonlinear structure formation in the presence of massive neutrinos and to study their effects on cosmological observables, which is needed to unlock the full potential of surveys like DESI and Euclid for neutrino science. Many approximate methods exist to incorporate neutrino effects in simulations, of which \cite{fidler19,zennaro19,partmann20,chen21,heuschling22} are some recent examples. Methods that solve for the neutrino and dark matter perturbations self-consistently fall roughly into three categories: grid-based methods actively solve evolution equations on the grid \cite{brandbyge09,viel10,hannestad11,archidiacono16,banerjee16,dakin17,inman20,yoshikawa20}, linear methods use transfer functions computed with an Einstein-Boltzmann code \cite{ali_haimoud12,tram19,chen21b}, and particle-based methods sample the phase-space distribution with tracers \cite{davis92,klypin92,ma94,brandbyge08,viel10,emberson17,castorina15,banerjee16,banerjee18,villaescusa_navarro20,elbers21}. While particle methods are uniquely suited to follow nonlinear neutrino clustering at late times, they typically disagree with linear theory in the neutrino component at early times, in part due to the way that initial conditions are handled and in part due to shot noise. The purpose of this paper is to address these shortcomings and to demonstrate how particle methods can be used to obtain accurate results at all times.

\begin{figure*}
	\normalsize
	\centering
	\subfloat{
		\includegraphics{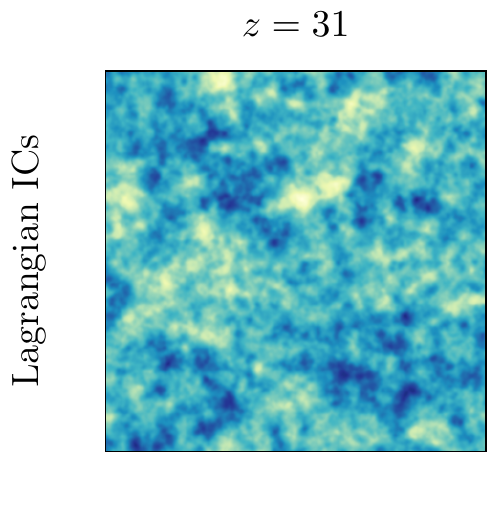}
	}\hspace{-3em}
	\subfloat{
		\includegraphics{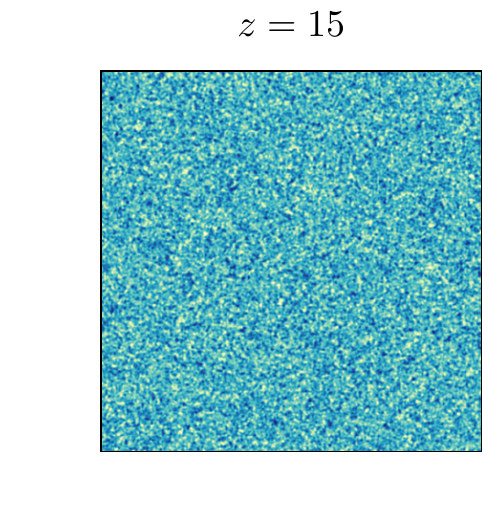}
	}\hspace{-3em}
	\subfloat{
		\includegraphics{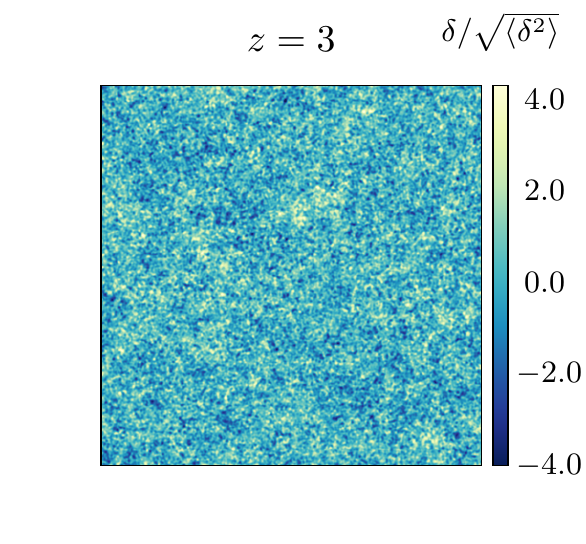}
	}\\\vspace{-5.5ex}
	\subfloat{
		\includegraphics{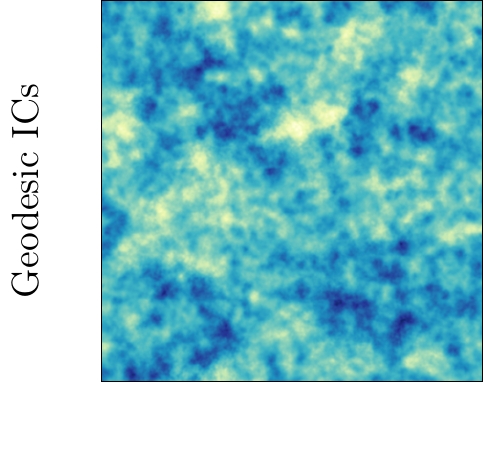}
	}\hspace{-3em}
	\subfloat{
		\includegraphics{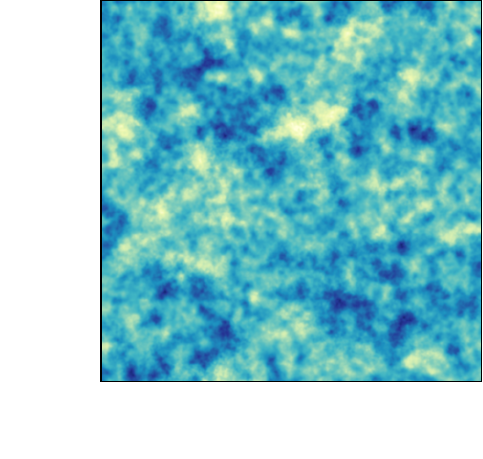}
	}\hspace{-3em}
	\subfloat{
		\includegraphics{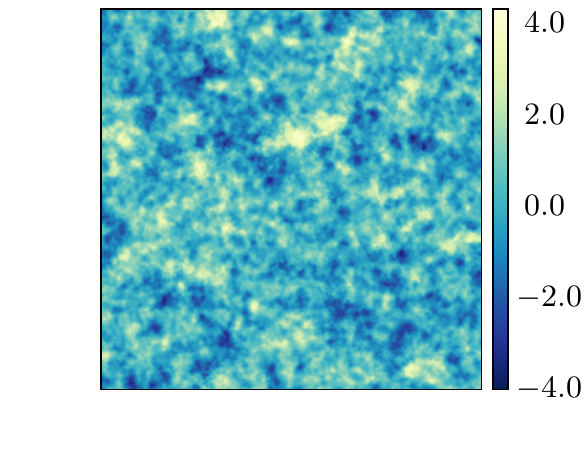}
	}
	\caption{Neutrino density slices from a $3.2\text{ Gpc}$ cube for $z\in\{31,15,3\}$. In the top row, neutrino particles were set up at $z=31$ using first-order Lagrangian perturbation theory (1LPT) and then integrated forward. The initial perturbations are immediately washed out and structure is only recovered over time. In the bottom row, the neutrino density field is faithfully reproduced at all times using geodesic integration together with the $\delta f$ method \cite{elbers21}.}
	\label{fig:density_plots}
\end{figure*}

Particle initial conditions for $N$-body simulations are commonly set up with Lagrangian perturbation theory (LPT). This works very well for baryons and cold dark matter, even in the presence of neutrinos \cite{wright17,aviles20,elbers22}. However, standard methods fail for the neutrino fluid itself. The free-streaming behaviour is usually implemented in an ad-hoc manner by drawing a random thermal velocity from the homogeneous Fermi-Dirac distribution and assigning it to the neutrino particles \cite{davis92,klypin92}. This is typically combined with first-order Lagrangian perturbation theory (1LPT), more commonly known as the Zel'dovich approximation \cite{zeldovich70}, in which particle displacements and velocities are proportional to one another: $\mathbf{v} = aHf\bm{\psi}$, where $f$ is the linear growth rate and $a$ the scale factor. It is easy to see that these steps are inconsistent. The result is illustrated in the top row of Fig.~\ref{fig:density_plots}. Even though the displacement field, $\bm{\psi}$, can be chosen to reproduce the density field at the initial time, the imprinted density perturbations are wiped out by random motions after only a few steps. A better approach, already proposed by \cite{ma94} and used recently by \cite{adamek17}, is to integrate neutrinos along geodesics from high redshift, $z=10^9$, down to the starting redshift of the simulation using metric perturbations obtained from an Einstein-Boltzmann code\footnote{Another solution could be to extend LPT to fluids with non-negligible velocity dispersion \cite{morita01,tatekawa02}.}. This, however, does nothing to address the issue of shot noise, which is particularly problematic at early times. We recently proposed the $\delta f$ method as a way of minimising shot noise in neutrino simulations \cite{elbers21}, inspired by similar efforts in plasma physics \cite{parker93,dimits93,aydemir94} and stellar dynamics \cite{merritt87,leeuwin93}. Here, we will show how these methods can be combined to produce accurate density fields from the very beginning of the simulation, as shown in the bottom panels of Fig.~\ref{fig:density_plots}. To facilitate this approach for large simulations, we have made our \FastDF{}\footnote{Fast Distribution Function; all codes available via \url{https://willemelbers.com/neutrino_ic_codes/}.} code publicly available, and integrated it into the higher-order initial conditions generator \monofonIC{}, along with other neutrino extensions \cite{michaux20,hahn20,elbers22b}.

The remainder of the paper is structured as follows. We will first describe our methods in \S2. We then derive the required equations of motion directly from the geodesic equation in \S3 and briefly remark on the Lagrangian derivation that was used previously. In \S4, we present numerical results, comparing the proposed method with linear fluid calculations and standard methods, and evaluating the impact of the equations of motions. Finally, we discuss the implications for simulations in \S5.

\section{Methods}

Throughout this paper, we work in Newtonian gauge with a metric given by
\begin{align}
\mathrm{d}s^2 = a^2(\tau)\left[-(1+2\psi(\mathbf{x},\tau))\mathrm{d}\tau^2 + (1-2\phi(\mathbf{x},\tau))\delta_{ij}\mathrm{d}x^i\mathrm{d}x^j\right], \label{eq:metric}
\end{align}

\noindent
where $\tau$ is conformal time and we consider only scalar metric perturbations: $\phi$ and $\psi$. Let $U_\mu=\mathrm{d}x_\mu/\sqrt{-\mathrm{d}s^2}$ be the 4-velocity and $P_\mu=mU_\mu$ the 4-momentum of a massive neutrino particle. The physical momentum measured by a cosmological observer is 
\begin{align}
    p=\sqrt{g_{ij}P^iP^j}.
\end{align}

\noindent
We define the comoving momentum as $q=ap$ and let $q^i = q_i = q\hat{n}_i$, where the unit vector $\hat{n}_i = P_i/P$ with $P^2=\delta^{ij} P_iP_j$. Finally, we define the energy as $\epsilon=\sqrt{q^2+m^2a^2}$. Our aim is to sample particles from the neutrino phase-space distribution,
\begin{align}
f(\mathbf{x},\mathbf{q},\tau) = \bar{f}(q)\left[1+\Psi(\mathbf{x},\mathbf{q},\tau)\right],\label{eq:def1}
\end{align}

\noindent
where $\bar{f}(q)=(1+\exp(q/T))^{-1}$ is the homogeneous Fermi-Dirac distribution and $T=\SI{1.95}{\K}$ the present-day neutrino temperature. In terms of $f$, the energy density is
\begin{align}
\rho(\mathbf{x},\tau) &= a^{-4}\int\mathrm{d}^3q\;\epsilon\,f(\mathbf{x},\mathbf{q},\tau) \label{eq:rho_def}\\
&= \bar{\rho}(\tau)\left[1+\delta(\mathbf{x},\tau)\right].
\end{align}

\noindent
The evolution of $f$ is governed by the collisionless Boltzmann equation
\begin{align}
    \frac{\partial f}{\partial\tau} + \frac{\mathrm{d} x^i}{\mathrm{d}\tau}\frac{\partial f}{\partial x^i} + \frac{\mathrm{d}q^i}{\mathrm{d}\tau}\frac{\partial f}{\partial q^i} = 0. \label{eq:boltz}
\end{align}

\noindent
At linear order in the metric perturbations, solutions can be found by decomposing $\Psi$ into a Legendre series in Fourier space \cite{ma95}:
\begin{align}
    \Psi(\mathbf{k},\hat{n},q,\tau) = \sum_{\ell=0}^\infty (-i)^\ell (2\ell+1)\Psi_\ell(\mathbf{k},q,\tau)P_\ell(\hat{k}\cdot\hat{n}). \label{eq:def2}
\end{align}

\noindent
The Boltzmann equation \eqref{eq:boltz} then becomes an infinite tower of equations in $\Psi_\ell$, which is usually truncated at some high $\ell_{\rm{max}}$ using an algebraic Ansatz. We solve this system with \CLASS{} \cite{lesgourgues11,lesgourgues11b}. To obtain very accurate results, we turn off the default neutrino fluid approximation and use $N=100$ momentum bins and an integration tolerance of $10^{-12}$. In terms of $\Psi_\ell$, the energy density and flux perturbations can then be written as\footnote{Recall that $\int_{0}^\pi P_\ell(\cos\theta)P_m(\cos\theta)\sin\theta\,\mathrm{d}\theta=\delta_{\ell m} 2/(2\ell +1)$.}
\begin{align}
\bar{\rho}(\tau){\delta}(\mathbf{k},\tau) &= a^{-4}\int\mathrm{d}^3q\;\epsilon\,\bar{f}(q)\Psi_0(\mathbf{k},q,\tau), \label{eq:delta}\\
\left(\bar{\rho}(\tau)+\bar{P}(\tau)\right){\theta}(\mathbf{k},\tau) &= a^{-4}\int\mathrm{d}^3q\;qk\,\bar{f}(q)\Psi_1(\mathbf{k},q,\tau), \label{eq:theta}
\end{align}

\noindent
where $\bar{\rho}$ and $\bar{P}$ are the background density and pressure.

\subsection{Initial conditions}

To sample particles from the full perturbed phase-space distribution \eqref{eq:def1}, taking into account the non-trivial correlations between $\mathbf{x}$ and $\mathbf{q}$, we integrate particles along geodesics from high redshift. We begin shortly after decoupling at $z=10^9$, when all modes of interest are outside the horizon and the neutrino phase-space distribution can be described in closed form, although in practice a slightly lower redshift would suffice. To recover the correct super-horizon evolution, we account for the initial monopole and dipole temperature perturbations. At early times on super-horizon scales, the first two moments of the distribution function are $\delta=-2\psi$ and $\theta=\tfrac{1}{2}k^2\tau\psi$ \cite{ma95}. From (\ref{eq:delta}--\ref{eq:theta}), we find
\begin{align}
    \Psi_0 = -\frac{\delta}{\nu}\frac{\mathrm{d}\log\bar{f}}{\mathrm{d}\log q},  \;\,\;\,\;\,\;\,\;\,\;\,\;\,   \Psi_1 &= -\frac{\omega\epsilon\theta}{qk\nu}\frac{\mathrm{d}\log\bar{f}}{\mathrm{d}\log q}, \label{eq:Psi_relations}
\end{align}

\noindent
where $\nu \equiv \mathrm{d}\log\bar{\rho}/\mathrm{d} \log T= 4$ and $\omega\equiv1+w=(\bar{\rho}+\bar{P})/\bar{\rho}=4/3$. It follow that
\begin{align}
    f(\mathbf{x},\mathbf{q},\tau) &= \bar{f}\left(q \left[1 - \frac{\delta}{\nu} - \frac{\omega\epsilon\hat{q}}{\nu q}\cdot\nabla\left(\nabla^{-2}\theta\right)\right]\right). \label{eq:pre_int}
\end{align}

\noindent
Particle positions are sampled uniformly in the periodic simulation volume. We then apply the initial perturbations by sampling momenta from the unperturbed Fermi-Dirac distribution, $\bar{f}$, and rescaling the $i$th component of $\mathbf{q}$:
\begin{align}
    q_i \to q_i\left[1 + \frac{\delta}{\nu} + \frac{\omega\epsilon}{\nu q_i}\nabla_i\left(\nabla^{-2}\theta\right)\right]. \label{eq:transform}
\end{align}

\noindent
After setting up these ``pre-initial'' conditions, neutrinos are integrated using relativistic equations motion, derived in the next section. These depend on the scalar potentials, $\phi(\mathbf{x},\tau)$ and $\psi(\mathbf{x},\tau)$, whose transfer functions are computed with \CLASS. The integration is done with the \textsc{C}-code \FastDF, which we make publicly available. Since the metric is computed in linear theory beforehand, each neutrino is completely independent, in principle allowing the code to be perfectly parallel. However, a large fraction of the computational expense is due to the potential grids, which can be shared if the particles are synchronised. To exploit this, \FastDF{} supports parallelisation through both \textsc{openmp} and \textsc{mpi}. The latter is also used to facilitate parallel data output through \textsc{hdf}5. Further gains in speed are made by realising that the metric perturbations are constant during pure radiation and pure matter domination. We therefore compute the potential fields only when the fractional change in the transfer functions exceeds $1\%$ and linearly interpolate between these super-steps. This significantly reduces the required number of Fourier transforms and has a negligible impact on the accuracy.

\subsection{The $\delta f$ method}

To handle particle shot noise, which is of particular concern at early times, we use the $\delta f$ method \cite{elbers21}. This is a variance reduction technique in which the phase-space distribution is decomposed as
\begin{align}
	f(\mathbf{x},\mathbf{q},\tau) = \bar{f}(q) + \delta f(\mathbf{x},\mathbf{q},\tau).
\end{align}

\noindent
In contrast to the usual approach, only the perturbation $\delta f$ is estimated from the particles. The density integral \eqref{eq:rho_def} is then decomposed into a smooth background, $\bar{\rho}(\mathbf{x},\tau)$, and a sum over simulation particles:
\begin{align}
	\rho(\mathbf{x},\tau) &\cong \bar{\rho}(\mathbf{x},\tau) + \frac{M}{N}\sum_{k=1}^N w_k\epsilon_k\, W(\mathbf{x}-\mathbf{x}_k), \label{eq:empirical_rho}
\end{align}

\noindent
where $M$ is a normalisation factor, $W(\mathbf{x})$ a smoothing kernel, $\epsilon_k$ the energy and $w_k$ a statistical weight for particle $k$ given by
\begin{align}
    w_k=\frac{\delta f(\mathbf{x}_k,\mathbf{q}_k,\tau)}{f(\mathbf{x}_k,\mathbf{q}_k,\tau)}.
\end{align}

\noindent
The weights are simple to compute in practice. Conservation of phase-space density along geodesics implies that $f(\mathbf{x}_k,\mathbf{q}_k,\tau) = \bar{f}(p_k)$ with $p_k$ the initially sampled (unperturbed) value for particle $k$ at $z=10^9$. At any later point, we obtain $\delta f(\mathbf{x}_k,\mathbf{q}_k,\tau) = \bar{f}(p_k) - \bar{f}(q_k)$. The method similarly extends to other phase-space statistics, such as the momentum density.

\section{Equations of motion}

We will derive the relativistic equations of motion starting directly from the geodesic equation and then comment on the differences with \cite{ma94,adamek17}.

\subsection{Geodesic derivation}\label{sec:geod_deriv}

To derive equations of motion in terms of $x^i$ and $q_i$, we begin with the geodesic equation $\nabla_P P=0$. Its components read
\begin{align}
	P^\nu\frac{\mathrm{d}P^i}{\mathrm{d}x^\nu} = - \Gamma^i_{\mu\nu}P^\mu P^\nu. \label{eq:geodesic_master}
\end{align}

\noindent
To first order, the Christoffel symbols $\Gamma^i_{\mu\nu}$ are
\begin{align}
	\begin{split}
    \Gamma^i_{00} &= \partial_i\psi,\\
    \Gamma^i_{j0} &= \delta^i_j(aH - \dot{\phi}),\\
    \Gamma^i_{jk} &= -2\partial_{(j}\phi\delta_{k)i} + \partial_i\phi\delta_{jk}.
	\end{split}
\end{align}

\noindent
Furthermore, using $q^2=a^2g_{ij}P^iP^j$ and $m^2 = -g_{\mu\nu}P^\mu P^\nu$, we express the momentum components in terms of the energy $\epsilon=\sqrt{q^2+m^2a^2}$ and the comoving 3-momentum $q_i$:
\begin{align}
	\begin{split}
	P^0 &= a^{-2}(1-\psi)\epsilon, \\
	P^i &= a^{-2}(1+\phi)q^i.
	\end{split} \label{eq:Pdef}
\end{align}

\noindent
The left-hand side of \eqref{eq:geodesic_master} consists of two terms, the first being
\begin{align}
    P^0\frac{\mathrm{d}P^i}{\mathrm{d}\tau} &= a^{-4}(1-\psi)\epsilon\left(-2aHq^i(1+\phi) + \dot{\phi}q^i + (1+\phi)\frac{\mathrm{d}q^i}{\mathrm{d}\tau}\right), \label{eq:geodesic_a}
\end{align}

\noindent
whereas the second is simply
\begin{align}
    P^j\frac{\mathrm{d}P^i}{\mathrm{d}x^j} &= a^{-4}(1+\phi)q^j\partial_j\phi q^i. \label{eq:geodesic_b}
\end{align}

\noindent
The right-hand side of \eqref{eq:geodesic_master} consists of three terms that can be written as
\begin{align}
    \Gamma^i_{\mu\nu}P^\mu P^\nu &= \Gamma^i_{00}a^{-4}(1-2\psi)\epsilon^2 + 2\Gamma^i_{0j}a^{-4}(1-\psi+\phi)q^j\epsilon + \Gamma^i_{jk}a^{-4}(1+2\phi)q^jq^k \\
	\begin{split}
    &= \partial_i\psi a^{-4}(1-2\psi)\epsilon^2 + \delta^i_j(2aH - 2\dot{\phi})a^{-4}(1-\psi+\phi)q^j\epsilon \\
    &\;\;\;\;\;+ \left(-2\partial_{(j}\phi\delta_{k)i} + \partial_i\phi\delta_{jk}\right)a^{-4}(1+2\phi)q^jq^k.
	\end{split}\label{eq:geodesic_c}
\end{align}

\noindent
Using \cref{eq:geodesic_a,eq:geodesic_b,eq:geodesic_c} in the geodesic equation \eqref{eq:geodesic_master} and dividing by $a^{-4}\epsilon(1+\phi-\psi)$, we finally obtain the acceleration
\begin{align}
	\frac{\mathrm{d}q_i}{\mathrm{d}\tau} = -\epsilon\partial_i\psi - \frac{q^2}{\epsilon}\partial_{i}\phi + \frac{1}{\epsilon}q_iq^j\partial_j\phi + q_i\dot{\phi}. \label{eq:releq}
\end{align}

\noindent
From \eqref{eq:Pdef}, we also obtain
\begin{align}
    \frac{\mathrm{d}x^i}{\mathrm{d}\tau} = \frac{q^i}{\epsilon}(1+\phi+\psi). \label{eq:geodesic}
\end{align}

\noindent
Eqs.~\eqref{eq:releq} and \eqref{eq:geodesic} are the desired equations of motion. These have a different form from those used previously by \cite{ma94,adamek17}. This is due to the choice of independent variables, as will be discussed in the next section.

\begin{figure}
	\normalsize
	\centering
	\includegraphics{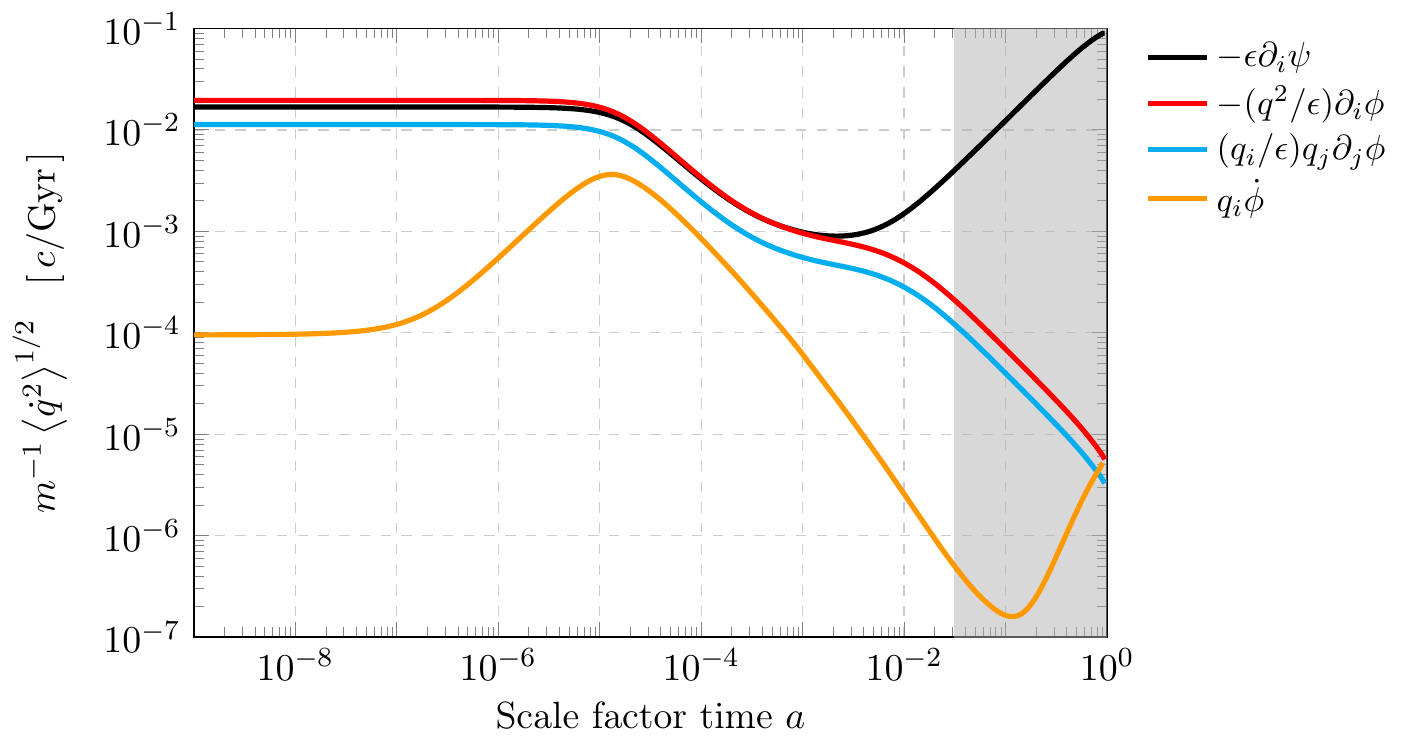}
	\caption{Contributions to the particle acceleration over time. The Newtonian acceleration, $-\epsilon\partial_i\psi$, dominates for $z\leq31$ (shaded), but the relativistic terms are relevant at early times.}
	\label{fig:eom_terms}
\end{figure}

\subsection{Lagrangian derivation}\label{sec:lagr_deriv}

The Lagrangian derivation\footnote{This use of `Lagrangian' should not be confused with references to Lagrangian perturbation theory or Lagrangian methods elsewhere.} of \cite{ma94} uses the same metric \eqref{eq:metric}, while \cite{adamek17} also include vector and tensor perturbations. Rather than working directly with the geodesic equation \eqref{eq:geodesic_master}, they start with the action:
\begin{align}
    S = \int\mathrm{d}\tau L = -m\int\sqrt{-\mathrm{d}s^2}. \label{eq:exact_Lagrangian}
\end{align}

\noindent
Expanding the Lagrangian $L$ to first order in the metric perturbations yields
\begin{align}
	L &= -m a \sqrt{1-u^2}\left[1+\frac{\psi+u^2\phi}{1-u^2}\right], \label{eq:approx_Lagrangian}
\end{align}

\noindent
where $u^i = \mathrm{d}x^i/\mathrm{d}\tau$ and $u^2=\delta_{ij}u^iu^j$. Observe that the second term inside the square brackets of \eqref{eq:approx_Lagrangian} diverges in the relativistic limit $u\to1$, so attention must be paid to the radius of convergence for fast particles. Proceeding from \eqref{eq:approx_Lagrangian}, the conjugate momentum variable to $x^i$ is found by differentiating the Lagrangian with respect to $u^i$:
\begin{align}
	P_i = \frac{\partial L}{\partial u^i} = \frac{m a u_i}{\sqrt{1-u^2}}\left(1 - 2\phi - \frac{\psi+u^2\phi}{1-u^2}\right). \label{eq:lagr_q_def}
\end{align}

\noindent
We note that \cite{ma94} here use the symbol $q_i$ for $P_i$, but stress that this conjugate momentum variable is in fact related to the comoving 3-momentum $q_i$ by a factor of $q_i/P_i=(1+\phi)$. The Euler-Lagrange equation gives
\begin{align}
	\frac{\mathrm{d}P_i}{\mathrm{d}\tau} = \frac{\partial L}{\partial x^i} = -\epsilon\partial_i\psi - \frac{P^2}{\epsilon}\partial_i\phi, \label{eq:releq_lag}
\end{align}

\noindent
where $P^2=P_iP_j\delta^{ij}$ and we used that $\epsilon=ma/\sqrt{1-u^2}$ to zeroth order. Meanwhile, inserting $u^i \propto (1 + f \psi + g \phi)$ into \eqref{eq:lagr_q_def} and solving for $f$ and $g$ gives
\begin{align}
	\frac{\mathrm{d}x^i}{\mathrm{d}\tau} = u^i = \frac{P_j\delta^{ij}}{\epsilon}\left[1 + \psi +\left(2-\frac{P^2}{\epsilon^2}\right)\phi\right]. \label{eq:geodesic_lag}
\end{align}

\noindent
The velocity corrections are small, so let us restrict attention to the acceleration equations \eqref{eq:releq} in terms of $q$ and \eqref{eq:releq_lag} in terms of $P$. Both equations contain the usual Newtonian acceleration $-\epsilon\partial_i\psi$ and a post-Newtonian term $-q^2/\epsilon\partial_i\phi$ or $-P^2/\epsilon\partial_i\phi$. However, the geodesic version \eqref{eq:releq} has two additional terms: another quadratic term $q_iq^j/\epsilon\partial_j\phi$ and a time-derivative or Sachs-Wolfe term $q_i\dot{\phi}$. These differences can be traced to the use of different momentum variables: the comoving 3-momentum $q_i$ in \S\ref{sec:geod_deriv} and the spatial part of the 4-momentum $P_i$ in \S\ref{sec:lagr_deriv}. The two quantities differ by a factor of $(1+\phi)$, which after insertion into \eqref{eq:releq_lag} yields the time-derivative term $q_i\dot{\phi}$. The quadratic term, meanwhile, arises in the geodesic derivation from the $P^iP^\mu_{,i}$ term of $\nabla_PP=0$. This quantity vanishes in the Lagrangian derivation, where the position $x^i$ and its conjugate momentum variable $P_i$ are independent. However, the term is generally non-zero when $q_i$ and $x^i$ are taken as independent instead. The question remains which choice of momentum variable is more suitable for neutrinos in $N$-body simulations. The advantage of $q_i$ is that it is a physical quantity, eliminating the dependence on metric perturbations when evaluating $\bar{f}(q)$. Since this is a necessary step for neutrino simulations, particularly when using the $\delta f$ method, we opt for the parametrisation in terms of $q$.

It is worth asking whether the relativistic corrections are needed in practice. In Fig.~\ref{fig:eom_terms}, we show the root mean square of the four terms of \eqref{eq:releq} between $z=10^9$ and $z=0$, for a $\SI{0.1}{\eV}$ neutrino. As expected, the acceleration is dominated by the Newtonian term (black) at late times. However, the relativistic corrections are non-negligible for $z>31$. Notably, the quadratic terms (red and blue) are always of the same order of magnitude and one should not be neglected if the other is included. Finally, the time-derivative term (yellow) is negligible during pure radiation or pure matter domination, but becomes relevant outside these r\'egimes. While the relativistic terms are clearly needed for generating initial conditions, they are less relevant for $N$-body simulations that are started sufficiently late, as will be discussed in \S5.

\subsection{Symplectic integration}\label{sec:symplectic}

Symplectic integrators explicitly conserve phase-space density\footnote{A linear map $J\colon\mathbb{R}^{2d}\to\mathbb{R}^{2d}$ is symplectic if $J^\mathrm{T}\Omega J=\Omega$ for $\Omega=\left(\begin{smallmatrix}0&I\\-I&0\end{smallmatrix}\right)$, with $I=I_d$ the $d\times d$ identity matrix. A differential map $f\colon U\to\mathbb{R}^{2d}$, with $U\subset\mathbb{R}^{2d}$ open, is symplectic if the Jacobian matrix $J$ of $f$ is everywhere symplectic. Conservation of phase-space density follows from $\det(J)=\det(\Omega)=1$.} and reduce the build-up of errors, which makes them suitable for $N$-body problems \cite{saha92,quinn97}. For \FastDF, we follow the simple strategy proposed in Appendix D of \cite{elbers21} and use separable equations of motion that closely approximate the relativistic form, yet admit a straightforward symplectic discretization, but see also Appendix A of \cite{adamek17} for a scheme involving a predictor-corrector step. Concretely, we approximate equations (\ref{eq:releq}--\ref{eq:geodesic}) with:
 \begin{align}
     \frac{\mathrm{d}\mathbf{q}}{\mathrm{d}\tau} &= -\epsilon_0\nabla\psi - \frac{q_0^2}{\epsilon_0}\nabla\phi + \frac{1}{\epsilon_0}\mathbf{q}_0 \left[\mathbf{q}_0\cdot\nabla\phi\right] + \mathbf{q}_0\dot{\phi}, \label{eq:releq_approx}\\
     \frac{\mathrm{d}x^i}{\mathrm{d}\tau} &= \frac{q^i}{\epsilon}, \label{eq:geodesic_approx}
 \end{align}

 \noindent
 where $\mathbf{q}_0 = \mathbf{q}(z=10^9)$ and $\epsilon_0=\sqrt{q_0^2+m^2a^2}$. Eq.~\eqref{eq:releq_approx} is a good approximation because $q_0\ll ma$ whenever $q$ deviates much from $q_0$: for slow particles at late times, while \eqref{eq:geodesic_approx} neglects the first-order term $\rvert\phi+\psi\rvert\ll1$. A leapfrog discretization of these equations is
\begin{align}
     \mathbf{q}_{k+\frac{1}{2}} &= \mathbf{q}_{k} + \Delta\tau_{k}^{k+\frac{1}{2}}\Big[-\epsilon_0\nabla\psi_k - \frac{q_0^2}{\epsilon_0}\nabla\phi_k + \frac{1}{\epsilon_0}\mathbf{q}_0 \left[\mathbf{q}_0\cdot\nabla\phi_k\right] + \mathbf{q}_0\dot{\phi}_k\Big],\\
     \mathbf{x}_{k+1} &= \mathbf{x}_{k} + \Delta\tau_{k}^{k+1}\frac{\mathbf{q}_{k+\frac{1}{2}}}{\sqrt{q_{k+\frac{1}{2}}^2 + m^2a^2}}, \label{eq:discrete_drift}\\
     \mathbf{q}_{k+1} &= \mathbf{q}_{k+\frac{1}{2}} + \Delta\tau_{k+\frac{1}{2}}^{k+1}\Big[-\epsilon_0\nabla\psi_{k+1} - \frac{q_0^2}{\epsilon_0}\nabla\phi_{k+1}+ \frac{1}{\epsilon_0}\mathbf{q}_0 \left[\mathbf{q}_0\cdot\nabla\phi_{k+1}\right]+ \mathbf{q}_0\dot{\phi}_{k+1}\Big],
\end{align}

\noindent
where $\psi_k=\psi(\mathbf{x}_k,a_k)$ and similarly for $\phi$. As is common in cosmological simulations, we use a constant step size $\Delta\log a$ and find the corresponding conformal time steps to be
\begin{align}
\Delta\tau_k^{\ell} &= \int_{\log a_k}^{\log a_\ell}\frac{\mathrm{d}\log a}{aH(a)}.
\end{align}

\noindent
We observe that $(\partial \mathbf{x}_{k+1}/\partial\mathbf{x}_k)(\partial \mathbf{q}_{k+1}/\partial\mathbf{q}_k)=I_d + (\partial \mathbf{x}_{k+1}/\partial\mathbf{q}_k)(\partial \mathbf{q}_{k+1}/\partial\mathbf{x}_k)$, which ensures sympecticity. To verify the validity of (\ref{eq:releq_approx}--\ref{eq:geodesic_approx}), we also implemented a non-symplectic leapfrog scheme based directly on (\ref{eq:releq}--\ref{eq:geodesic}) and found relative differences in the resulting power spectra of order $10^{-5}$, well below other sources of error.

\section{Results}

\begin{figure}
	\normalsize
	\centering
	\includegraphics{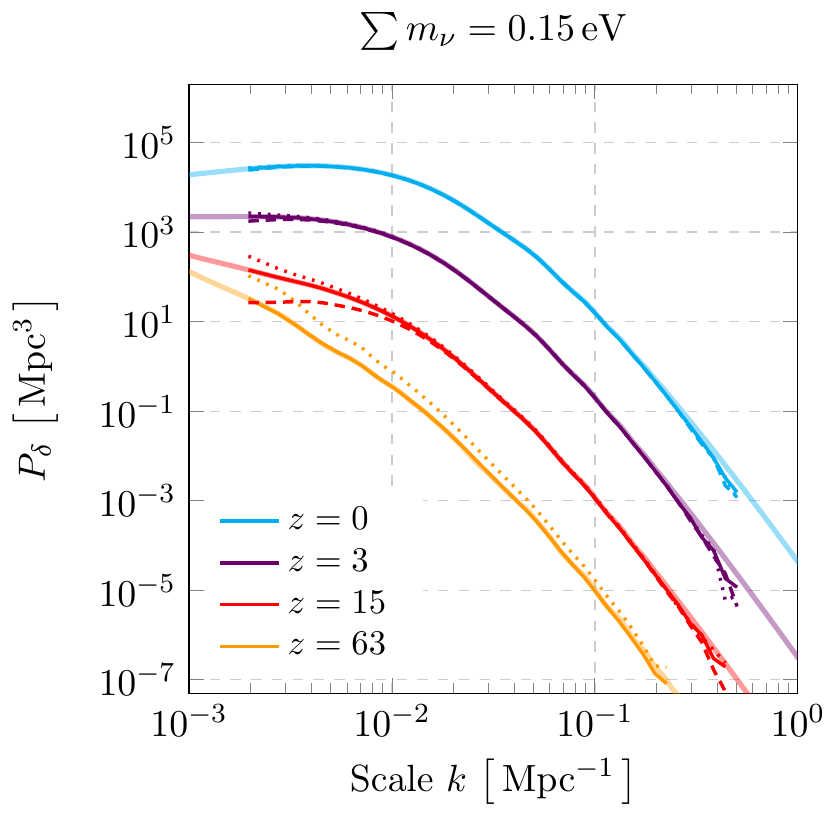}\hspace{-1em}
    \includegraphics{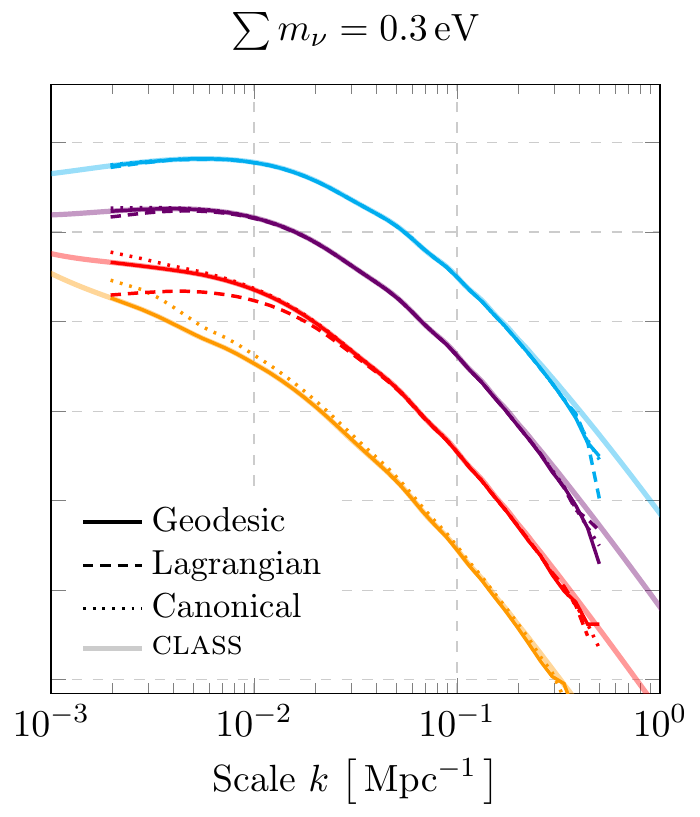}
	\caption{The linear neutrino density power spectrum at various redshifts computed from $800^3$ particles in an $L=3.2$ Gpc cube for $\sum m_\nu = \SI{0.15}{\eV}$ and $\sum m_\nu = \SI{0.3}{\eV}$. Particles were set up with Eq.~\eqref{eq:transform} at $z=10^9$ or with Lagrangian ICs at $z=31$ and subsequently evolved forward using linear metric perturbations. We also show the effect of substituting the canonical momentum $P$ for the comoving momentum $q$ in the Fermi-Dirac function. The spectra are compared with the linear fluid prediction from \CLASS. There is no line for the Lagrangian ICs at $z=63$.}
	\label{fig:neutrino_power_spectra}
\end{figure}

\begin{figure}
	\normalsize
	\centering
	\includegraphics{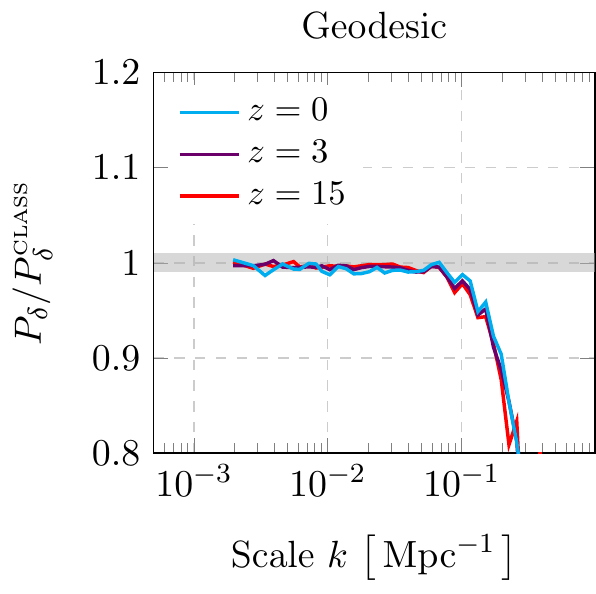}\hspace{-0.75em}
    \includegraphics{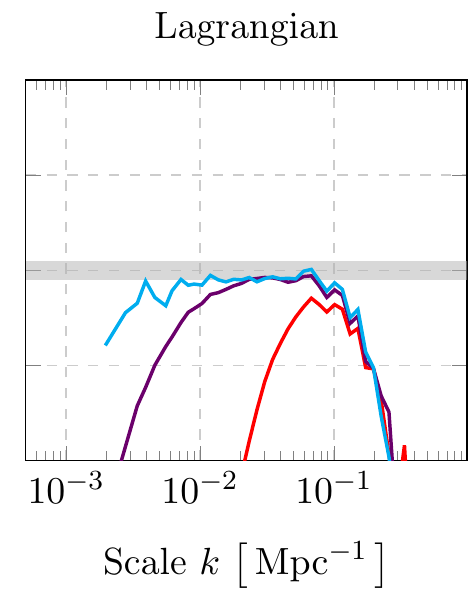}\hspace{-0.75em}
    \includegraphics{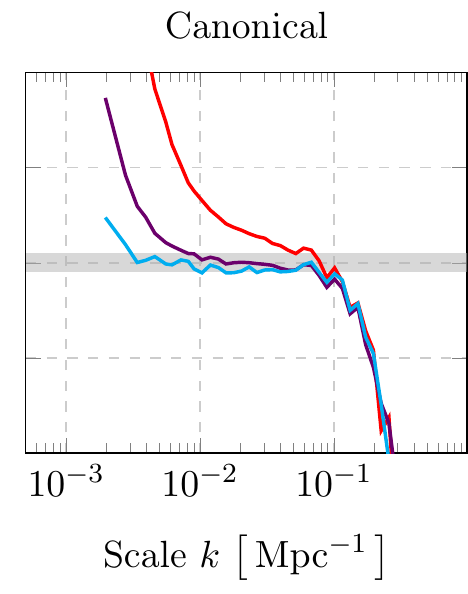}
	\caption{Ratios of the linear neutrino density power spectrum at various redshifts computed from $800^3$ particles in an $L=3.2$ Gpc cube for $\sum m_\nu = \SI{0.3}{\eV}$, relative to the linear fluid prediction from \CLASS. Particles were set up with Eq.~\eqref{eq:transform} at $z=10^9$ (left) or with Lagrangian ICs at $z=31$ (middle) and subsequently evolved forward using linear metric perturbations. We also show the effect of substituting the canonical momentum $P$ for the comoving momentum $q$ in the Fermi-Dirac function (right). The shaded area is $1\%$.}
	\label{fig:errors}
\end{figure}

We set up $800^3$ particles in a periodic volume with side length $L=\SI{3.2}{Gpc}$, using \eqref{eq:transform} to generate pre-initial conditions at $z=10^9$. For comparison, particles are also set up with first-order Lagrangian ICs at $z=31$\footnote{This is the fiducial starting redshift for neutrino ICs in \cite{elbers22}. Usually, $z=31$ is too late for accurate first-order ICs, but this is not true for neutrinos. Moreover, all calculations are linear in this paper.}. We consider two degenerate models with $\sum m_\nu=\SI{0.15}{\eV}$ $(f_\nu=0.11)$ and $\sum m_\nu=\SI{0.3}{\eV}$ ($f_\nu=0.023$). Fixed initial conditions are used to facilitate comparison with linear theory on large scales \cite{angulo16} and the $\delta f$ method is used in each case to suppress shot noise. First, we show the neutrino density power spectrum evaluated at various redshifts in Fig.~\ref{fig:neutrino_power_spectra}. Power spectra are computed by dividing the neutrino ensemble in half and taking the cross-spectrum, which eliminates the constant shot noise plateau on small scales \cite{inman15}. Note that we compute the power spectrum of the energy density, as expressed in \eqref{eq:empirical_rho}, as opposed to the mass density. The results are compared with the linear fluid calculations from \CLASS{}. We remind the reader that particles were integrated using linear metric perturbations, which should result in perfect agreement with \CLASS{}. We see that this is indeed the case with the geodesic approach, while the power is significantly underestimated for the runs with Lagrangian ICs, recovering only over time. We also show the effect of using the alternative equations (\ref{eq:releq_lag}--\ref{eq:geodesic_lag}), essentially substituting the canonical momentum $P$ for the comoving momentum $q$ without accounting for the relative factor $(1+\phi)$. In this case, the power spectrum is overestimated. In both cases, the errors are largest at early times, but persist on large scales down to $z=0$. This can be seen more clearly in Fig.~\ref{fig:errors}, where we show the ratios relative to \CLASS{} for $\sum m_\nu=\SI{0.3}{\eV}$. Using the geodesic method, we obtain $1\%$-agreement independent of redshift, while the other methods result in significant errors on all scales. At $z=0$, a ($-8\%$, $+5\%$) error remains at $k=\SI{2e-3}{\per\Mpc}$ when using Lagrangian ICs or when substituting the canonical momentum $P$ for $q$, respectively. In a full $N$-body simulation, this lack or excess of neutrino clustering would cause a back-reaction, resulting in still larger errors and contaminating the dark matter and baryon components.

\begin{figure}
	\normalsize
	\centering
	\includegraphics{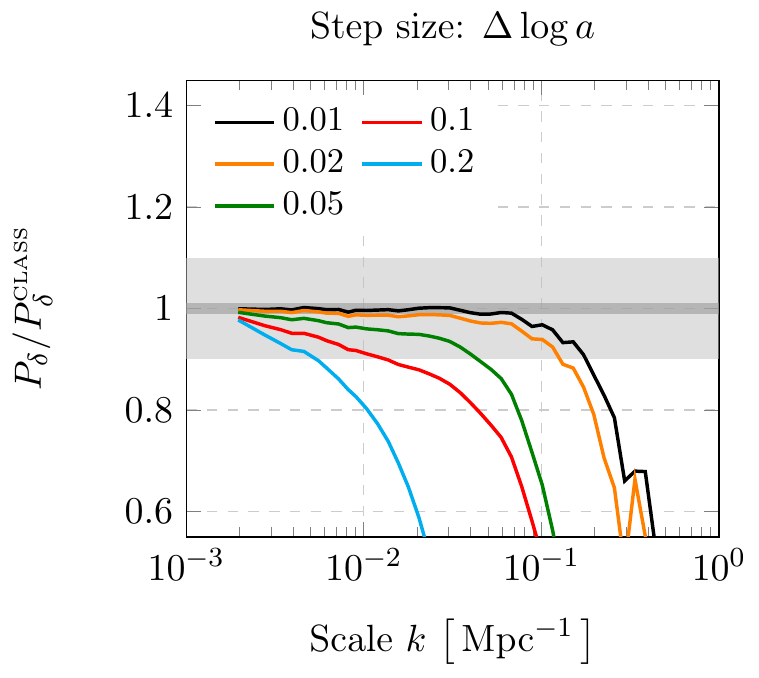}
    \includegraphics{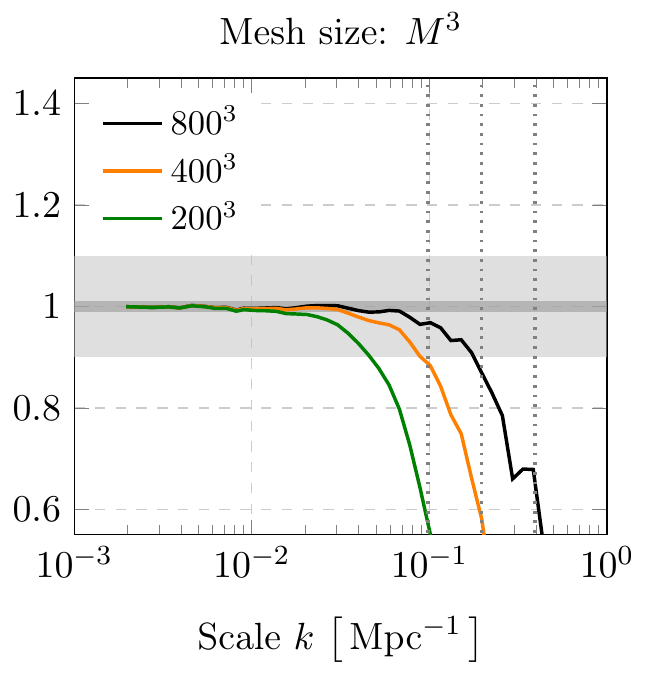}
	\caption{Impact of the step size (left) and the mesh size (right) on the neutrino density power spectrum at $z=31$, computed from $800^3$ particles in an $L=3.2$ Gpc cube for $\sum m_\nu = 0.3$ eV. The spectra are compared with the linear fluid prediction from \CLASS. The shaded areas are $1\%$ (dark) and $10\%$ (light). The vertical dotted lines on the right represent half the Nyquist frequency $k_N=\pi M/L$.}
	\label{fig:param_impact}
\end{figure}

In all cases, the power diminishes relative to \CLASS{} beyond $k=\SI{0.1}{\per\Mpc}$. This is due to the limited resolution of the runs. The precision and speed of \FastDF{} are mainly determined by two parameters: the step size $\Delta\log a$ and the size $M$ of the mesh on which the potentials are calculated. A third parameter, the interpolation order used when computing forces, chosen from $r=1$ or $r=2$, has a small effect on the accuracy. We show the impact of the first two parameters on the neutrino power spectrum at $z=31$ for $\sum m_\nu=\SI{0.3}{\eV}$ in Fig.~\ref{fig:param_impact}. For the main results in this paper, we used $\Delta\log a=0.01$ together with $M=800$, resulting in $1\%$-agreement with the fluid calculations up to $k=\SI{0.07}{\per\Mpc}$. However, errors decrease quickly on small scales in an $N$-body simulation once neutrinos become non-relativistic, so obtaining agreement on large scales is most important. For many applications, the parameters could therefore be relaxed to enable more rapid realisations of the neutrino distribution function.

\begin{figure}
	\normalsize
	\centering
	\includegraphics{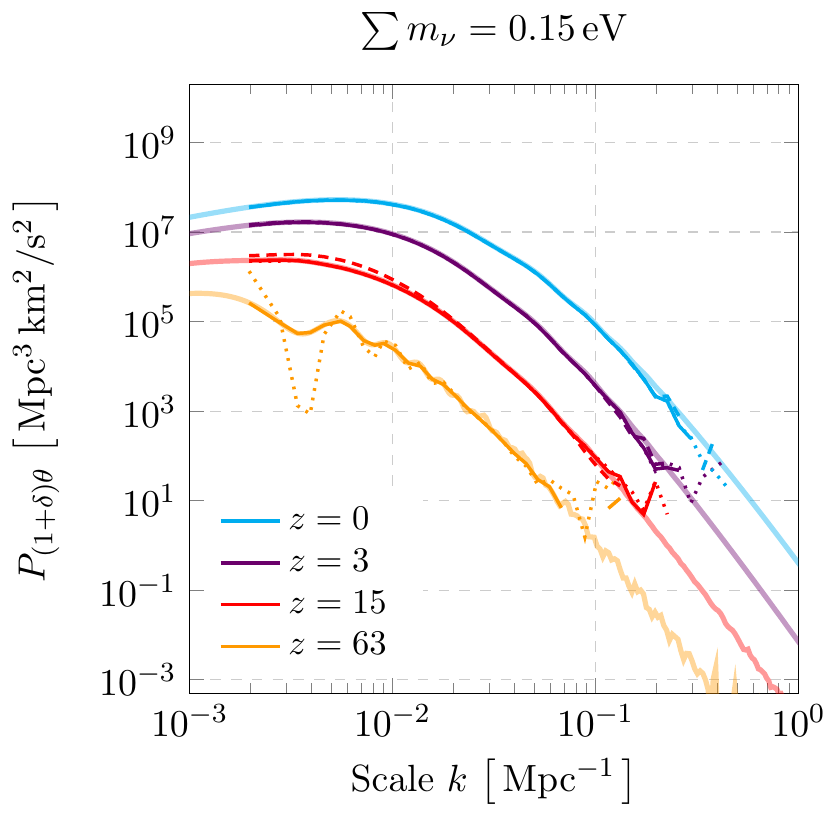}\hspace{-1em}
    \includegraphics{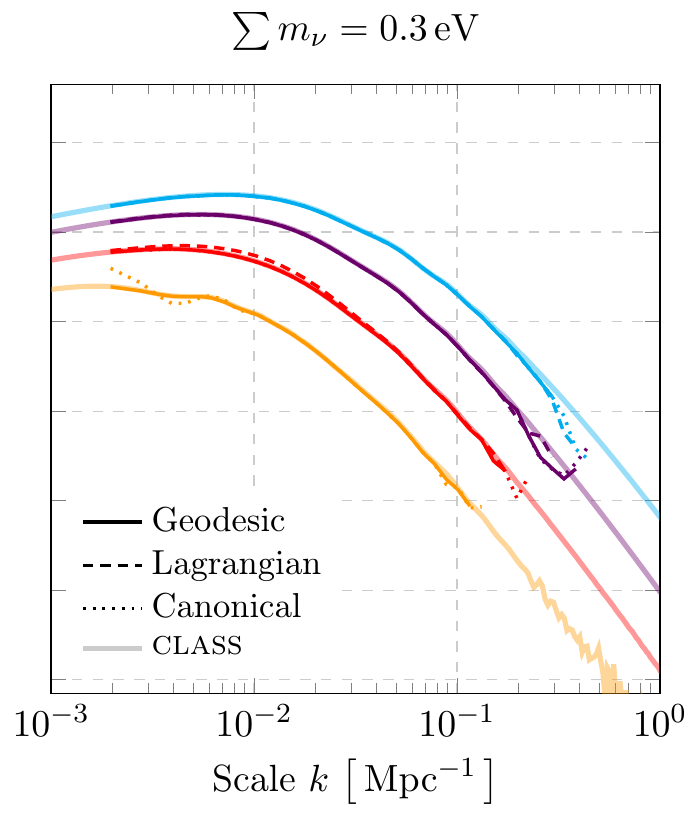}
	\caption{The linear neutrino momentum power spectrum at various redshifts computed from $800^3$ particles in an $L=3.2$ Gpc cube for $\sum m_\nu = \SI{0.15}{\eV}$ and $\sum m_\nu = \SI{0.3}{\eV}$. Particles were evolved in the linearly perturbed spacetime. The spectra are compared with the linear fluid prediction from \CLASS, which shows some scatter on small scales at early times. There is no line for the Lagrangian ICs at $z=63$.}
	\label{fig:neutrino_momentum_power_spectra}
\end{figure}

To demonstrate that we can also reproduce higher-order moments of the distribution function, we show the power spectrum of the momentum perturbation, $(1+\delta)\theta$, in Fig.~\ref{fig:neutrino_momentum_power_spectra}. Despite the extreme precision settings, a small scatter can be seen at large $k$ for the \CLASS{} results at $z\geq 15$, reflecting the difficulty of solving the Boltzmann hierarchy numerically on small scales. We once again obtain excellent agreement between the geodesic results and \CLASS{}, but find large errors at $z=63$ when using the canonical momentum, especially for the lighter neutrinos. Large errors are also apparent for the Lagrangian ICs at $z=15$. In contrast to the density power spectrum, however, these errors decrease quickly on large scales.

\section{Discussion}

The accurate treatment of massive neutrinos in cosmological $N$-body simulations, consistent with the demand of surveys like DESI and Euclid for percent-level accurate modelling of large-scale structure observables, also calls for accurate neutrino initial conditions (ICs). In this paper, we have shown that by integrating neutrino particles from high redshift, it is possible to obtain $1\%$-agreement with linear fluid calculations, even at early times. To achieve this level of agreement, suitable pre-initial conditions must be generated at sufficiently early times, the equations of motion must remain valid in the relativistic limit, and shot noise must be significantly suppressed. We addressed these requirements by providing a closed form expression for the super-horizon perturbations of the Fermi-Dirac distribution $f(\mathbf{x},\mathbf{q},\tau)$, by expressing the geodesic equation in terms of $\mathbf{q}$, and by using the $\delta f$ method to limit shot noise. We also used fixed ICs \cite{angulo16} to limit cosmic variance, which allowed a detailed comparison between linear particle and fluid methods.

When these conditions are not satisfied, significant errors in the neutrino component occur on large scales. For neutrino particles used in $N$-body simulations, this error causes a back-reaction in the dark matter and baryon components. Simulations that use Lagrangian or unperturbed ICs together with an ad-hoc momentum sampled from the homogeneous Fermi-Dirac distribution underestimate the clustering of neutrinos, leading to errors of a few percent on large scales at $z=0$. These errors get progressively worse at higher redshifts. Neutrino clustering recovers over time, beginning on small scales where errors are less apparent. Some simulations use hybrid methods (e.g. \cite{brandbyge09,bird18,adamek17}), transitioning from a linear or grid-based method at early times to a particle method at late times. This would mitigate the back-reaction arising from these errors. Nevertheless, we have demonstrated that a transition of this sort is not necessary if suitable ICs are used and shot noise is addressed.

These results have further implications for neutrino particles in $N$-body simulations. Aside from the ICs, some codes also use relativistic equations of motion for the neutrino particles in the simulation itself. For ordinary Newtonian simulations, \cite{adamek17} proposed using special relativistic equations of motion with Newtonian gravity. These can be obtained from (\ref{eq:releq_lag}--\ref{eq:geodesic_lag}) by assuming that $\rvert\phi\rvert\ll 1$ and $\phi\approx\psi$:
\begin{align}
	\frac{\mathrm{d}x^i}{\mathrm{d}\tau} &= \frac{P^i}{\sqrt{P^2+m^2a^2}},\\
	\frac{\mathrm{d}P_i}{\mathrm{d}\tau} &= -\frac{2P^2 + m^2a^2}{\sqrt{P^2+m^2a^2}}\partial_i\psi. \label{eq:p_eq}
\end{align}

\noindent
Based on (\ref{eq:releq}--\ref{eq:geodesic}) and Fig.~\ref{fig:eom_terms}, we instead propose the simpler form
\begin{align}
	\frac{\mathrm{d}x^i}{\mathrm{d}\tau} &= \frac{q^i}{\sqrt{q^2+m^2a^2}},\\
	\frac{\mathrm{d}q_i}{\mathrm{d}\tau} &= -\sqrt{q^2+m^2a^2}\,\partial_i\psi. \label{eq:q_eq}
\end{align}

\noindent
The velocities have the same form: including the Lorentz factor is crucial for sub-light neutrino speeds and physical free-streaming lengths. However, the accelerations are different due to the choice of momentum variable. By expressing the equations in terms of the physical quantity $q$, the potential $\phi$ need not be evaluated when computing the $\delta f$ weights. Even so, the corrections to the acceleration matter less in the time frame where Newtonian simulations are used to best effect ($z\ll 10^2$). Simply using the Newtonian acceleration, $\dot{q}_i = -m\partial_i\psi$, together with the special relativistic velocity equation therefore seems to be a reasonable alternative with the benefit of having a straightforward symplectic discretization. Let us remark finally on the choice of gauge. While Newtonian gauge is convenient for the geodesic integration, recent years have seen the introduction of gauges more naturally suited for cosmological $N$-body simulations. A popular choice is $N$-body gauge \cite{fidler15,fidler17}, in which the spatial metric perturbation is constant and traceless, such that the relativistic dark matter density coincides with that of the Newtonian simulation. Using \CLASS, it is possible to compute the shifts in density, $\Delta\delta(k)$, and energy flux, $\Delta\theta(k)$, from Newtonian to $N$-body gauge. Provided that the perturbations are small, the gauge transformation can then be applied to the neutrino ensemble in the same way as the pre-initial conditions, via \eqref{eq:transform}, since the higher-order moments $\Psi_\ell$ are gauge-invariant. This feature is available in \FastDF.

The main application of the described method is to set up accurate and consistent neutrino particle initial conditions for simulations. Another interesting application would be to integrate particles back along the line of sight from Earth to analyse the angular dependence of the local neutrino flux. Sampling the full phase-space distribution with particles may be advantageous if, for instance, non-trivial selections are of interest (e.g. neutrinos with momenta in a given interval that passed through halos in a particular mass range). If the metric perturbations are treated in linear theory, as in this paper, the method could provide a cross-check of linear calculations \cite{hannestad09,tully21}, while transitioning from an $N$-body simulation at late times would enable a fully nonlinear calculation. Another interesting extension would be to consider other massive thermal relics \cite{lesgourgues11b,banerjee22}.

\acknowledgments

I thank Carlos Frenk, Adrian Jenkins, Baojiu Li, and Silvia Pascoli for their comments and encouragement. I also thank Volker Springel, C\'esar Hern\'andez-Aguayo, and the members of the Euclid Neutrino Method Comparison Project for useful discussions. WE is supported by the Durham Prize Scholarship in Astroparticle Physics. This work was also supported by STFC Consolidated Grants for Astronomy at Durham ST/P000541/1 and ST/T000244/1. This work used the DiRAC@Durham facility managed by the Institute for Computational Cosmology on behalf of the STFC DiRAC HPC Facility (www.dirac.ac.uk). The equipment was funded by BEIS capital funding via STFC capital grants ST/K00042X/1, ST/P002293/1 and ST/R002371/1, Durham University and STFC operations grant ST/R000832/1. DiRAC is part of the National e-Infrastructure.

\appendix

\bibliographystyle{JHEP}
\bibliography{main}

\end{document}